# Gauge and emergent electromagnetic fields for moving magnetic topological solitons

K.Y. Guslienko,[1,2,a]

[1]*Depto. Física de Materiales, Universidad del País Vasco, UPV/EHU, 20018 San Sebastián, Spain*

[2]*IKERBASQUE, the Basque Foundation for Science, 48013 Bilbao, Spain*

We apply the general conception of non-Abelian gauge fields for description of magnetic soliton excitations. We show that the component of the gauge field along the soliton local magnetization (Abelian part of the gauge potential) determines dynamics of spin fluctuations over the soliton background in a ferromagnet. Assumption that the gauge field is a pure gauge allows calculating the gauge field components and finding simple expressions for the emergent electromagnet fields related to the soliton motion. The gauge field results in soliton-magnon interaction leading to renormalization of the soliton and magnon dynamics. The presented approach allows reaching more deep understanding of a relationship of the field theory and condensed matter magnetism.

[a]Electronic address: kostyantyn.gusliyenko@ehu.es

The gauge symmetry and gauge fields are one of the central concepts in modern physics. They are widely used in the classical and quantum field theory describing the non-linear topologically non-trivial field excitations in the form of monopoles, instantons, skyrmions etc. [1]. In condensed matter physics topological magnetic solitons [2] of the magnetization field in ferromagnets such as vortices [3], skyrmions etc. were also discovered. The inhomogeneous magnetization textures (vortices, skyrmions, domain walls, etc.) and their dynamics in restricted geometry are being studied extensively for the last 10 years. There is considerable interest nowadays to emergent electromagnetic fields which are generated by the magnetization textures and essentially influence on the conductivity electrons, magnetic soliton motions and spin waves [4-21].

To describe magnetic soliton excitations in a ferromagnet it is convenient to use a local coordinate frame related to the local instant soliton magnetization direction $\mathbf{M}(\mathbf{r},t)$. In this frame small fluctuations around the soliton background have the simplest form. The price of that is appearance of gauge fields, which one has to introduce to keep the Lagrangian and equations of motion independent on choice of the coordinate system. The gauge transformations are described by using some continuous groups of symmetry and are related to some charges and their conservation. The simplest example of the gauge fields is electromagnetic Abelian U(1) gauge field. In the case of the magnetic topological solitons those charges are topological charges. We introduce below the gauge fields for magnetic solitons and demonstrate how this conception is useful for description of the soliton dynamics and magnon (spin wave) dynamics on the soliton background. We consider gauge fields acting on magnetization vector in a ferromagnet. Therefore, these fields represented by square 3×3 matrices are non-Abelian and described by SO(3) Lie group of 3D-rotations [22]. We show below that projections of the gauge fields on local instant quantization axis have sense of electromagnetic fields of topological origin, or so called ¨emergent¨ electromagnetic fields. The conception of the SU(2) emergent fields was widely used to describe electric current interaction with an inhomogeneous magnetic texture.

Transformation to the local spin quantization axis in the kinetic part of an electronic Hamiltonian was first conducted by Korenman et al. [24]. Then, this approach was developed within the *s-d* exchange model of a conducting ferromagnet, see Ref. 6 and references therein. It essentially depends on assumption about adiabaticity of the response of the conductivity *s*-electrons to the magnetization texture (*d*-electrons) local variations for the dominating *s-d* interaction. Such approach was successfully applied to explain the giant magnetoresistance, spin transfer torque, topological Hall effect and other related effects. To consider magnetic soliton dynamics we do not need any assumptions about adiabatic response and non-adiabatic effects of the conductivity electrons because the magnetization fluctuations are considered by definition in a moving local frame related to the soliton background magnetization.

We consider magnetization field $\mathbf{M}(\mathbf{r},t)$ dynamics in a ferromagnet and write the corresponding Lagrangian via the magnetization field components and their derivatives with respect to space $\mathbf{r}$ and time coordinates. The Lagrangian should be invariant at arbitrary coordinate system rotations. Therefore, we introduce covariant derivatives $D_\mu = (\partial_\mu - A_\mu)$, where 3x3 matrices $A_\mu$ are components of a non-Abelian gauge potential (the index $\mu$ = 0, 1, 2, 3 denotes the time and space coordinates $x_\mu$ = *ct, x, y, z* and $\partial_\mu = \partial/\partial x_\mu$). Let us consider a rotaion of the laboratory Cartesian coordinate system *xyz* to a new coordinate sysytem *x´y´z´*, where *Oz´* axis is parallel to the soliton magnetization represented by the unit vector **m** ($\mathbf{m}(\mathbf{r},t) = \mathbf{M}(\mathbf{r},t)/M_s$, $M_s$ is the saturation magnetization, and $\mathbf{m}^2 = 1$), see Fig. 1. The passive rotation to the local quantization axis is defined as $\mathbf{m}' = R\mathbf{m}$, where *R* is the rotation matrix. One can show that the gauge potential $A_\mu$ is transformed to new coordinate sysytem as $A_\mu \rightarrow A'_\mu = R A_\mu R^{-1} + \partial_\mu R \cdot R^{-1}$. The first term is indefinite (we can put $A_\mu = 0$) and we concentrate our atention on the second ¨pure¨ gauge term $\hat{A}_\mu = \partial_\mu R \cdot R^{-1}$, which leads to numereous physical consequencies for the magnetization dynamics. The unitary matrix *R* can be represented in the standard exponential form $R = \exp(i\psi \mathbf{n} \cdot \mathbf{J})$, where $\psi$ is an angle of rotation around the axis directed along an unit vector $\mathbf{n}$, and $\mathbf{J}$ is the operator of angular momentum ($J=1$). Due its definition the matrix of the pure gauge potential $\hat{A}_\mu$ is antisymmetric and therefore can be represented by a dual vector $\mathbf{A}_\mu$. Direct calculation yields the expressions $\hat{A}_\mu = 2(\partial_\mu \mathbf{n} \otimes \mathbf{n} - \mathbf{n} \otimes \partial_\mu \mathbf{n})$ and $\mathbf{A}_\mu = 2(\partial_\mu \mathbf{n} \times \mathbf{n})$, where the symbol $\otimes$ marks Kronecker product of two vectors. However, we do not know the vector $\mathbf{n}$ and, therefore, it is more convenient to represent the gauge field $\hat{A}_\mu$ in terms of the local soliton magnetization $\mathbf{m}(\Theta, \Phi)$ defined by the spherical angles $\Theta, \Phi$ (Fig. 1). The identity $\hat{A}_\mu \mathbf{m} = \mathbf{A}_\mu \times \mathbf{m}$ holds for any vector $\mathbf{m}$ if we define the vector $\mathbf{A}_\mu$ using the decomposition $\hat{A}_\mu = (\mathbf{A}_\mu \cdot \mathbf{G}) = A_\mu^a G_a$, where $G_a = iJ_a$, $(G_a)_{bc} = \varepsilon_{abc}$, $a,b,c=x,y,z$ are the generators of the group SO(3) in the Cartesian coordinate representation [22]. For the pure gauge field $D_\mu \mathbf{m}' = (\partial_\mu - \hat{A}_\mu)\mathbf{m}' = (\partial_\mu - \partial_\mu R R^{-1})R\mathbf{m} \equiv 0$ or $\partial_\mu \mathbf{m} = \mathbf{A}_\mu \times \mathbf{m}$ for the vector $\mathbf{m}(\Theta, \Phi)$, which was used to construct the matrix $R$ in the transformation $\mathbf{m}' = R\mathbf{m}$. In our case, this is a ¨vacuum¨ or soliton background magnetization vector. The spin waves (SW) then correspond to excitation over this ¨vacuum¨ state. *I.e.*, we distinguish two subsystems: a magnetic soliton + magnons (spin waves) and express magnetization as a sum $\mathbf{m}_t = \mathbf{m}\sqrt{1 - \mathbf{m}_s^2} + \mathbf{m}_s$ of the soliton and SW orthogonal contributions, $\mathbf{m} \cdot \mathbf{m}_s = 0$. We consider SW magnetization $\mathbf{m}_s$ as a small perturbation of the moving soliton $\mathbf{m}$ background and calculate a mutual influence of the soliton and SW dynamics. The square root is equal to 1 in the case of linear magnon dynamics considered below. The covariant derivatives $D_\mu \mathbf{m}_s$ include a soliton-magnon interaction expressed via the gauge field potential $\mathbf{A}_\mu$. The components of $\mathbf{m}_s$ are the simplest in a moving coordinate frame $x'y'z'$, $\mathbf{m}'_s = (\vartheta, \sin\Theta \psi, 0)$, where the axis $Oz'$ is directed along the local direction of $\mathbf{m}$ defined by the spherical angles of $\mathbf{m}(\Theta,\Phi)$, and the angles $\vartheta,\psi$ are small deviations from the soliton $\mathbf{m}$ angles $(\Theta,\Phi)$.

Using the equations $D_\mu \mathbf{m} = 0$ and the conservation of the magnetization length $\partial_\mu \mathbf{m} \cdot \mathbf{m} = 0$ allows us to write immediately the gauge potential in an invariant form as a sum of the longitunal and transverse components with respect to the quantization axis $\mathbf{m}$ direction

$$\mathbf{A}_\mu = \mathbf{m} \times \partial_\mu \mathbf{m} + (\mathbf{A}_\mu \cdot \mathbf{m})\mathbf{m} . \qquad (1)$$

Eq. (1) is the decomposition of $\mathbf{A}_\mu$ in the basis of three mutually perpendicular vectors $\mathbf{m}$, $\partial_\mu \mathbf{m}$ and $\mathbf{m} \times \partial_\mu \mathbf{m}$ in the isospin space. The covariant derivative $D_\mu \mathbf{m}_t$ is reduced to $D_\mu \mathbf{m}_s = \partial_\mu \mathbf{m}_s - A^e_\mu \mathbf{m} \times \mathbf{m}_s$, where $A^e_\mu = (\mathbf{A}_\mu \cdot \mathbf{m})$. Therefore, only the longitudinal component of the gauge potential $A^e_\mu$ contributes to the magnon Lagrangian. The component $A^e_\mu$ plays a special role in the soliton and spin wave dynamics. Sometimes, the component $A^e_\mu$ (an Abelian part of the non-Abelian gauge field potential $\mathbf{A}_\mu$ [23]) is called as the potential of an ¨emergent¨ U(1) electromagnetic field. The emergent field should be added to the Maxwell U(1) electromagnetic field considering magnetization dynamics.

The question is how to find the components of $\mathbf{A}_\mu$. We can use the orthogonal rotation matrix representation as $R(\mathbf{n},\psi) = \exp(i\psi \mathbf{n} \cdot \mathbf{J})$ or $R(\alpha,\beta,\gamma) = \exp(i\gamma J_z)\exp(i\beta J_y)\exp(i\alpha J_z)$ via the Euler angles $\alpha,\beta,\gamma$. The latter parametrization yields [8, 24]

$$A^x_\mu = \sin\gamma \partial_\mu \beta - \sin\beta \cos\gamma \partial_\mu \alpha, \quad A^y_\mu = \cos\gamma \partial_\mu \beta + \sin\beta \sin\gamma \partial_\mu \alpha, \quad A^z_\mu = \partial_\mu \gamma + \cos\beta \partial_\mu \alpha. \quad (2)$$

The rotation $R(\Phi,\Theta,0)$ corresponds to transformation to the spherical coordinate system in magnetization space $(\hat{\mathbf{m}}, \hat{\mathbf{\Theta}}, \hat{\mathbf{\Phi}})$ with $\mathbf{m} \| \hat{\mathbf{m}}$. The unit vector $\mathbf{m}$ is defined as $\mathbf{m} = (\sin\Theta\cos\Phi, \sin\Theta\sin\Phi, \cos\Theta)$ in the laboratory coordinate system or as $\mathbf{m}' = (0,0,1)$ in the local coordinate system $x'y'z'$. Here the unit vectors in isospin space are $\hat{\mathbf{z}}' = \hat{\mathbf{m}}$, $\hat{\mathbf{x}}' = \hat{\boldsymbol{\Theta}}$, $\hat{\mathbf{y}}' = \hat{\boldsymbol{\Phi}}$. Many authors were restricted themselves by the rotation $R(\Phi, \Theta, 0)$ considering that the third rotation on the angle $\gamma$ is an arbitrary and has no influence on the Lagrangian and magnetization equations of motion. Due to the expression for $D_\mu \mathbf{m}_s$ one has to use not z-component $A_\mu^z$ in the laboratory sysytem [13, 14, 15, 24, 25] but rather the longitudinal $\hat{\mathbf{m}}$-component of the gauge potential $\mathbf{A}_\mu$. The longitudinal component of the gauge potential is

$$A_\mu^e(\gamma) = (\mathbf{A}_\mu \cdot \mathbf{m}) = \sin(\gamma+\Phi)\sin\Theta \partial_\mu \Theta + [\cos^2\Theta - \sin^2\Theta\cos(\gamma+\Phi)]\partial_\mu \Phi + \cos\Theta \partial_\mu \gamma. \quad (3)$$

The SO(3) gauge field $\hat{A}_\mu$ is non-Abelian and we cannot build the field tensor similar to Maxwell U(1) Abelian electrodynamics as $F_{\mu\nu} = \partial_\mu \hat{A}_\nu - \partial_\nu \hat{A}_\mu$. However, one can construct a tensor of the background soliton Yang-Mills field [1] as a gauge covariant curl (transforming at rotations as $F_{\mu\nu} \to F'_{\mu\nu} = R F_{\mu\nu} R^{-1}$)

$$F_{\mu\nu} = \partial_\mu \hat{A}_\nu - \partial_\nu \hat{A}_\mu - [\hat{A}_\mu, \hat{A}_\nu], \quad (4)$$

which can be written in the vector form

$$\mathbf{F}_{\mu\nu} = \partial_\mu \mathbf{A}_\nu - \partial_\nu \mathbf{A}_\mu + [\mathbf{A}_\mu \times \mathbf{A}_\nu], \quad (4')$$

where $F_{\mu\nu} = (\mathbf{F}_{\mu\nu} \cdot \mathbf{G}) = F_{\mu\nu}^a G_a$ is the dual vector representation.

Let us consider the field tensor component $(\mathbf{F}_{\mu\nu} \cdot \mathbf{m}) = \partial_\mu A_\nu^e - \partial_\nu A_\mu^e + \mathbf{m} \cdot [\mathbf{A}_\mu \times \mathbf{A}_\nu]$. Using Eq. (1) this component can be calculated as $(\mathbf{F}_{\mu\nu} \cdot \mathbf{m}) = F_{\mu\nu}^e + \mathbf{m} \cdot [\partial_\nu \mathbf{m} \times \partial_\mu \mathbf{m}]$, where $F_{\mu\nu}^{e} = \partial_{\mu} A_{\nu}^{e} - \partial_{\nu} A_{\mu}^{e}$ is the emergent electromagnetic field tensor. This is a particular case of the Faddeev-Niemi field tensor [26] for the case of absence of the auxiliary scalar fields $\rho, \sigma$. From the other side, it is well known that the field tensor components $F_{\mu\nu} \equiv 0$ and the energy density $w(\hat{A}_{\mu}) \propto Tr(F_{\mu\nu} F_{\mu\nu}) = 0$ for the pure gauge field defined as $\hat{A}_{\mu} = \partial_{\mu} R \cdot R^{-1}$. Therefore, the condition $F_{\mu\nu}^{e} = \mathbf{m} \cdot [\partial_{\mu} \mathbf{m} \times \partial_{\nu} \mathbf{m}]$ holds and defines the longitudinal gauge field component $A_{\mu}^{e}$:

$$\partial_{\mu} A_{\nu}^{e} - \partial_{\nu} A_{\mu}^{e} = \sin\Theta (\partial_{\mu} \Theta \partial_{\nu} \Phi - \partial_{\mu} \Phi \partial_{\nu} \Theta). \tag{5}$$

Accounting Eq. (3) this condition leads to choice of the definite value of the Euler angle $\gamma = \pi - \Phi$ and, therefore, $A_{\mu}^{e} = (1 - \cos\Theta) \partial_{\mu} \Phi$. The choices $\gamma = 0$ [12, 14, 15, 19, 24, 27] and $\gamma = -\Phi$ used in Ref. [16] do not satisfy to Eq. (5) and should be disregarded. Using the relation of the Euler angles and the Rodriguez parameters $(\mathbf{n}, \psi)$ one can find the set $\mathbf{n} = \left(\sin\frac{\Theta}{2}\cos\Phi, \sin\frac{\Theta}{2}\sin\Phi, \cos\frac{\Theta}{2}\right)$, $\psi = \pi$. The rotation matrix can be written in the following simple symmetric form $R(\mathbf{n}, \pi) = \exp(i\pi \mathbf{n} \cdot \mathbf{J}) = 2\mathbf{n} \otimes \mathbf{n} - 1$. It describes rotation by the angle $\pi$ around the axis $\mathbf{n} = (\hat{\mathbf{z}} + \mathbf{m})/|\hat{\mathbf{z}} + \mathbf{m}|$ directed in the $\hat{\mathbf{z}} - \hat{\mathbf{m}}$ plane at the angle $\Theta/2$ with respect to $\hat{\mathbf{z}}$-axis. The matrix $R(\mathbf{n}, \pi) = 2\mathbf{n} \otimes \mathbf{n} - 1$ was used in Ref. [13] for SO(3) and the SU(2) transformation matrix $U = -iR(\mathbf{n}, \pi)$ was used in Refs. [4-7, 28] to diagonalize the *s-d* exchange Hamiltonian, $U^{+}(\mathbf{n} \cdot \boldsymbol{\sigma}) U = \sigma_{z}$, where $\mathbf{J} = \boldsymbol{\sigma}/2$ is the spin operator for *s*=1/2. Only the diagonal part of the gauge potential is accounted in the electron-magnetic texture interaction theories that corresponds to projection to the electronic sub-bands with the spin $s = \pm 1/2$ [11, 28, 29].

The U(1) field tensor $F_{\mu\nu}^{e} = \partial_{\mu} A_{\nu}^{e} - \partial_{\nu} A_{\mu}^{e}$ allows to define the emergent magnetic and electric fields $B_{i} = -\varepsilon_{ijk} F_{jk}^{e}/2$, $E_{i} = F_{i0}^{e}$, and skyrmion number (topological charge) $S = \int d^3x \varepsilon_{\mu\nu} F^e_{\mu\nu} / 4\pi$. The emergent fields $\mathbf{E} = \nabla A^e_0 - \partial_0 \mathbf{A}^e$, $\mathbf{B} = \nabla \times \mathbf{A}^e$ can be expressed via the $A^e_\mu$ potential components $(A^e_0, \mathbf{A}^e)$ as in the Maxwell electrodynamics ($V^e = -A^e_0$ is the emergent electric field potential).

The covariant derivatives $D_\mu \mathbf{m}_s = \partial_\mu \mathbf{m}_s - A^e_\mu \mathbf{m} \times \mathbf{m}_s$, $D_\mu \mathbf{m} = 0$ and the emergent field potential $A^e_\mu = (1 - \cos\Theta) \partial_\mu \Phi$ allows to construct the Lagrangian $L(D_\mu \mathbf{m}_s, \mathbf{m}, \mathbf{m}_s)$, which is invariant with respect to the local rotations $R$.

To describe magnetization dynamics we use the Lagrangian

$$\Lambda = \int d^3\mathbf{r} L(\mathbf{r}, t), \qquad L = \mathbf{C}(\mathbf{m}) \cdot D_0 \mathbf{m} - w(\mathbf{m}, D_a \mathbf{m}), \qquad (6)$$

where $\mathbf{C}(\mathbf{m}) = \Im (1 + \mathbf{m} \cdot \hat{\mathbf{z}})^{-1} [\hat{\mathbf{z}} \times \mathbf{m}]$ is the vector potential of the Dirac monopole in isotropic space, $\Im = M_s / \gamma$ is the spin angular momentum density, $w$ is the magnetic energy density $w = A(D_a \mathbf{m})^2 + w_d$, $A$ is the exchange stiffness, and $x_a = x, y, z$. The term $w_d$ includes all other energy contributions (magnetostatic, anisotropy, Zeeman, etc.), which do not contain the magnetization derivatives and the gauge fields. Note that the magnetostatic energy density $w_m = -M_s \mathbf{m} \cdot \mathbf{H}_m / 2$ includes the non-local magnetostatic field, $\mathbf{H}_m[\mathbf{m}]$, which is a functional of the magnetization $\mathbf{m}$. The potential $\mathbf{C}(\mathbf{m})$ allows the simple definition $A^e_\mu = \Im^{-1} \mathbf{C}(\mathbf{m}) \cdot \partial_\mu \mathbf{m}$ that relates the longitudinal and transverse parts of the gauge potential $\mathbf{A}_\mu$.

Using the substitution $\mathbf{m}_t = \mathbf{m}\sqrt{1 - \mathbf{m}_s^2} + \mathbf{m}_s$ and introducing the magnetic soliton center coordinate $\mathbf{X}$ via an appropriate ansatz $\mathbf{m}(\mathbf{r}, t) = \mathbf{m}[\mathbf{r}, \mathbf{X}(t)]$ the Lagrangian (6) can be written as

$$\Lambda = \Lambda[\mathbf{m}] + \int d^3\mathbf{r} [\mathbf{C}(\mathbf{m} + \mathbf{m}_s) \cdot D_0 \mathbf{m}_s - w(\mathbf{m}_s, D_a \mathbf{m}_s)], \qquad (7)$$

where $\Lambda[\mathbf{m}] = G_{ab}X_a\dot{X}_b/2 - W(\mathbf{X})$ is the soliton Lagrangian, $W(\mathbf{X}) = \int d^3\mathbf{r} w[\mathbf{m}(\mathbf{r},\mathbf{X})]$ is the soliton energy, and $G_{ab} = \Im\int d^3\mathbf{r}\mathbf{m}\cdot[\partial_{X_a}\mathbf{m}\times\partial_{X_b}\mathbf{m}]$ is the gyrocoupling tensor [30]. In the case of the rigid soliton motion $\mathbf{m}(\mathbf{r},t) = \mathbf{m}[\mathbf{r} - \mathbf{X}(t)]$, the gyrocoupling tensor $G_{ab} = \Im\int d^3\mathbf{r} F^e_{ab}$ is simply averaged emergent field tensor. The soliton-magnon interaction is solely determined by the covariant derivatives $D_\mu = (\partial_\mu - \hat{A}_\mu)$ in the Lagrangian (7). This is similar to the vacuum - material field interaction in the Yang-Mills field theory. The time derivatives in $\mathbf{A}_\mu$, $D_\mu$ were not considered in Refs. [8, 20] that did not allow to consider properly spin wave dynamics over the soliton background. The bilinear kinetic interaction term in Eq. (7) $L_{\text{int}} = \mathbf{C}(\mathbf{m})\cdot D_0\mathbf{m}_s = \Im\mathbf{m}_s\cdot\mathbf{A}_0$ is important for magnetization dynamics. However, it was overlooked in Ref. [9] describing magnons over the skyrmion background. Alternatively, the soliton-magnon interaction can be written as an effective Zeeman energy $L_{\text{int}} = M_s\mathbf{m}_s\cdot\mathbf{H}_g$, where $\mathbf{H}_g(\mathbf{m}) = c\mathbf{A}_0(\mathbf{m})/\gamma$ is the dynamic gyrofield acting on the spin wave magnetization [31].

If accounting the SW magnetization $\mathbf{m}'_s = (\vartheta, \sin\Theta\psi, 0)$ we introduce the variable $\Psi = \vartheta + i\sin\Theta\psi$ [25], then the magnon exchange energy can be expressed as $w_{ex}(\mathbf{m}_s) = A(D_a\mathbf{m}_s)^2 = A|(\nabla - i\mathbf{A}^e)\Psi|^2$, where $\mathbf{A}^e = (1-\cos\Theta)\nabla\Phi$. The spatial part of the Abelian gauge potential $\mathbf{A}^e$ in the quadratic approximation is $\mathbf{A}^e = (1-\cos\Theta_0)\nabla\Phi_0$, where $\Theta_0(\rho)$, $\Phi_0$ are the spherical angles of the magnetiztion $\mathbf{m}[\mathbf{r},0]$ of a static, cylindrically symmetric soliton. For particular case of 2D vortex and $\varphi-$ ($\rho-$) skyrmion accounting $\Phi_0 = \pi/2(0) + q\varphi$ we get explicitly the potential $\mathbf{A}^e = q(1-\cos\Theta_0)\hat{\varphi}/\rho$, and the emergent magnetic field $\mathbf{B} = \hat{\mathbf{z}}q\Theta'_0\sin\Theta_0/\rho$ [13, 25], where $(\rho,\varphi)$ are the cylindrical coordinates. The flux of this field $\int d^2\boldsymbol{\rho}B_z$ does not depend on the details of the soliton profile $\Theta_0(\rho)$ and is proportional to the skyrmion number $S = pq/2$, where $p = \cos\Theta_0(0)$.

In summary, introducing the pure gauge potential ($D_\mu \mathbf{m} = 0$) we derived the explicit expression for the non-Abelian SO(3) gauge field potential $\mathbf{A}_\mu$ in a ferromagnet, which corresponds to arbitrary local rotations in 3D-space. The longitudinal component of the potential along the soliton magnetization direction (the Abelian component $A_\mu^e = (\mathbf{A}_\mu \cdot \mathbf{m})$) enters the magnetization equations of motion of solitons and magnons determining emergent U(1) electromagnetic field. The magnetic component of the emergent field determines the soliton topological charge. The emergent electromagnetic field is a result of the non-Abelian character of the gauge field induced by the moving magnetic soliton and is determined by the gauge field self-interaction terms in Eq. (4). The gauge field leads to some interaction of the soliton background with spin waves represented by spatial and time derivatives of the soliton magnetization. The formulated approach can serve as a benchmark for any calculations of spin excitations in inhomogeneous magnetic textures.

**Acknowledgements**

The support by IKERBASQUE (the Basque Science Foundation) and the Spanish MINECO grant MAT2013-47078-C2-1-P is acknowledged.

# References

[1] A. Zee, *Quantum field theory in a nutshell* (Princeton Univ. Press, Princeton, 2010).

[2] A.M. Kosevich, B.A. Ivanov, and A.S. Kovalev, *Phys. Rep.* **194**, 117 (1990).

[3] K.Y. Guslienko, *J. Nanosci. Nanotechn.* **8**, 2745-2761 (2008).

[4] J. Shibata, Y. Nakatani, G. Tatara, H. Kohno, and Y. Otani, *Phys. Rev.* B **73**, 020402 (2006).

[5] G. Tatara, and H. Kohno, *Phys. Rev. Lett.* **92**, 086601 (2004).

[6] G. Tatara, H. Kohno, and J. Shibata, *Phys. Rep*. **468**, 213 (2008).

[7] M.B. Jalil, S.G. Tan, K. Eason, and J.F. Kong, *J. Appl. Phys.* **115**, 17D107 (2014).

[8] V. K. Dugaev, P. Bruno, B. Canals, and C. Lacroix, *Phys. Rev.* B **72**, 024456 (2005).

[9] C. Schuette, and M. Garst, *Phys.Rev.* B **90**, 094423 (2014).

[10] N. Nagaosa, and Y. Tokura, *Nat. Nanotech.* **8**, 899 (2013); N. Nagaosa, X.Z. Yu, Y. Tokura, Phil. Trans. R. Soc. **A 370**, 5806 (2012).

[11] P. Bruno, V.K. Dugaev, M. Taillefumier, *Phys. Rev. Lett.* **93**, 096806 (2004).

[12] Y. Yamane, J. Ieda, J. Ohe, S.E. Barnes, S. Maekawa, *J. Appl. Phys.* **109**, 07C735 (2011).

[13] K.A. van Hoogdalem, Y. Tserkovnyak and D. Loss, *Phys. Rev.* B **87**, 024402 (2013).

[14] A.A. Kovalev, Y. Tserkovnyak, *EPL* **97**, 67002 (2012).

[15] A.A. Kovalev, *Phys. Rev.* B **89**, 241101 (2014).

[16] G. Tatara, N. Nakabayashi, *J. Appl. Phys*. **115**, 172609 (2014).

[17] K.Y. Guslienko, G.N. Kakazei, J. Ding, X.M. Liu, and A.O. Adeyeye, *Sci. Rep.* **5**, 13881 (2015).

[18] F. Büttner, et al., *Nature Phys*. **11**, 225 (2015).

[19] K.Y. Guslienko, G.R. Aranda, and J. Gonzalez, *Phys. Rev*. B **81**, 014414 (2010).

[20] Y.-T. Oh, H. Lee, J.-H. Park, J.H. Han, *Phys. Rev*. B **91**, 104435 (2015).

[21] T. Schulz, R. Ritz, A. Bauer, M. Halder, M. Wagner, C. Franz, C. Pfleiderer, K. Everschor, M. Garst, and A. Rosch, *Nat. Phys*. **8**, 301 (2012).

[22] L.C. Biedenharn and J.D. Louck, *Angular momentum in quantum physics* (Addison-Wesley Publ. Co., Reading, Massachusetts, 1981), Ch. 2,3.

[23] G. ´t Hooft, *Nucl. Phys.* B **190**, 455 (1981).

[24] V. Korenman, J. L. Murray, and R. E. Prange, *Phys. Rev.* B **16**, 4032 (1977); **16**, 4048 (1977); **16**, 4058 (1977).

[25] D. D. Sheka, I. A. Yastremsky, B. A. Ivanov, G. M. Wysin, and F. G. Mertens, *Phys. Rev.* B **69**, 054429 (2004).

[26] L. Faddeev, and A.J. Niemi, *Phys. Rev. Lett.* **82**, 1624 (1999).

[27] J. Ohe, and S. Maekawa, *J. Appl. Phys.* **105**, 07C706 (2009).

[28] S. Zhang, and S. S.-L. Zhang, *Phys. Rev. Lett.* **102**, 086601 (2009).

[29] Ya. B. Bazaliy, B. A. Jones, and S.-C. Zhang, *Phys. Rev.* B **57**, R3213 (1998).

[30] A.A. Thiele, *Phys. Rev. Lett.* **30**, 230-233 (1973).

[31] K.Y. Guslienko, A.N. Slavin, V. Tiberkevich, and S.-K. Kim, *Phys. Rev. Lett.* **101**, 247203 (2008).

**The captions to figures**

Fig. 1. The laboratory Cartesian coordinate system *xyz* and local coordinate frame *x´y´z´* with the *z´*-axis oriented parallel to local instant soliton magnetization $\mathbf{m}(\Theta,\Phi)$, where $(\Theta,\Phi)$ are the magnetization spherical angles.

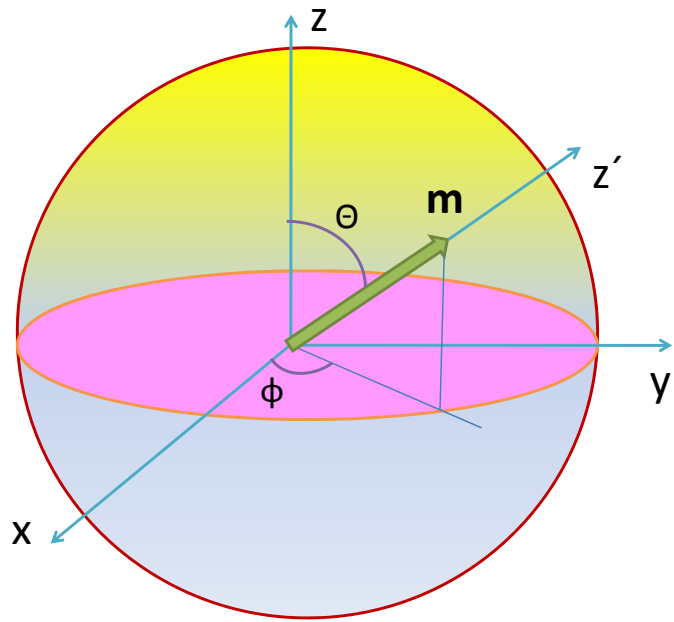